\documentclass[prd,aps,floats,epsfig,eqsecnum,nofootinbib]{revtex4}
\usepackage{amsmath,amssymb,verbatim,epsfig,psfig,graphicx,rotating}
\newcommand{\be}{\begin{equation}}
\newcommand{\ee}{\end{equation}}
\newcommand{\bea}{\begin{eqnarray}}
\newcommand{\eea}{\end{eqnarray}}
\begin{document}
\title{\bf
THE SELF-GRAVITATING GAS IN THE PRESENCE OF DARK ENERGY: MONTE-CARLO 
SIMULATIONS AND STABILITY ANALYSIS}
\author{ {\bf  H. J. de Vega$^{(a,b)}$, J. A. Siebert$^{(a)}$\\}}
\affiliation{$^{(a)}$Laboratoire de Physique Th\'eorique et Hautes Energies, \\
Universit\'e Paris VI et Paris VII,  Laboratoire Associ\'e au CNRS UMR 7589,
Tour 24, 5\`eme \'etage, \\ 4, Place Jussieu
75252 Paris cedex 05, France.}
\affiliation{$^{(b)}$Observatoire de Paris, LERMA.
Laboratoire Associ\'e au CNRS UMR 8112.
 \\61, Avenue de l'Observatoire, 75014 Paris, France.}
\begin{abstract}
The self-gravitating gas in the presence of a positive cosmological constant 
$ \Lambda $ is studied in thermal equilibrium by Monte Carlo simulations and by 
the mean field approach. We find excellent agreement between both approaches already 
for $ N = 1000 $ particles on a volume $V$ [The mean field is exact in the infinite 
$N$ limit]. The domain of stability of the gas is found to increase when the 
cosmological constant increases. The particle density is shown to be an increasing
(decreasing) function of the distance when the dark energy dominates over
self-gravity (and vice-versa).
We confirm the validity of the thermodynamic limit: $ N, V \to   \infty $ 
with $ N/V^{\frac13} $ and $\Lambda \; V^{\frac23} $ fixed.
In such dilute limit extensive thermodynamic quantities like energy, free energy, 
entropy turn  to be proportional to $N$.
We find that the gas is stable till the isothermal compressibility diverges.
Beyond this point the gas becomes a extremely dense object whose properties 
are studied by Monte Carlo. 
\end{abstract} 
\date{\today} 
\maketitle
\tableofcontents

\section{Introduction}
The self-gravitating gas in thermal equilibrium has been thoroughly
studied since many years\cite{gasn,gas2,Emdsg,Chandrsg,Ebertsg,Bonnorsg,
Antonovsg,Lyndsg,Jeans,BT}. As a
consequence of the long range attractive Newton force, the
selfgravitating gas admits a consistent thermodynamic limit
  $ N, V \to   \infty $ with $ \frac{N}{V^{\frac13}} $ fixed. In this limit,
  extensive thermodynamic quantities like energy, free energy, entropy
  are proportional to $N$\cite{gasn,gas2}.

In ref.\cite{gaslambda} we investigated how a cosmological constant
affects the properties of the non-relativistic self-gravitating gas
in thermal equilibrium by mean field methods. The mean field approximation
becomes exact in the limit when the number of particles becomes infinity.

In the present paper we study the stability properties of the self-gravitating
gas in the presence of a cosmological constant by  mean field and Monte-Carlo 
methods.
The use of Monte-Carlo simulations is particularly useful since they are
like real experiments and allow to unambiguously determine the stability
or instability of the system. 

We compute both by Monte Carlo and mean field methods several physical 
quantities: the equation of state, the particle density and the average 
particle distance for different values of the dark energy. An excellent
agreement between both approaches is found except very close to the 
collapse transition [see figs. 4-6]. The difference between both approaches
turns to be, as expected, of the order $ \frac1{N} $ where the number of 
particles $N$ we choose in the  Monte Carlo simulations was 
$ N = 1000 - 2000 $. The  slightly 
larger split between  Monte Carlo and mean field near the collapse is due to 
the fact that there the corrections
to the mean field become singular \cite{gasn,gas2}.

We find that the onset of instability in the canonical ensemble
coincides with the point where the isothermal compressibility diverges. 
At this point we find that the dimensionless parameter $ \zeta = 
\frac{G^3 \; m^6 N^2 \; P}{T^4} $ is maximal. ($ \zeta $ was introduced in 
ref. \cite{Lyndsg} for $ \Lambda = 0 $). We find that  the 
domain of stability of the gas increases for increasing cosmological
constant. The dark energy has an anti-gravity effect that disfavours the 
collapse pushing apart the particles.

In absence of cosmological constant the particle density $\rho(r)$ 
is a {\bf decreasing}
function of the distance in the self-gravitating gas (see for example 
\cite{gasn}). In the presence of a positive  $ \Lambda $ we find that 
 $\rho(r)$ {\bf decreases} with $r$ when the self-gravity dominates
over the dark energy. This happens for $ X \equiv \frac{2 \Lambda V}{m N} < 1$.
For $X>1$ the cosmological constant dominates over the self-gravity and
the the particle density turns to {\bf increase} with the distance.
This is a consequence of the repulsive character of the dark energy.

\medskip

The  Monte-Carlo study of the self-gravitating gas shows the validity of
 the dilute thermodynamic limit introduced in refs.\cite{gasn,gaslambda}.
Namely,  for $ N, V \to   \infty $ with $ \frac{N}{V^{\frac13}} $ and 
$\Lambda \; V^{\frac23} $ fixed, a consistent thermodynamic limit is reached
where extensive thermodynamic quantities like energy, free energy, entropy
are  shown to be proportional to $N$. 
Furthermore, the Monte-Carlo simulations allow us to study 
the condensed phase which turns to be an extremely dense body. 
The phase transition to collapse is found to be of zeroth order.

\medskip

The outline of the paper is as follows. Section II presents non-relativistic 
selfgravitating particles in the presence of the cosmological constant, 
their statistical mechanical treatment and the mean field approach.
In section III we analyze the stability of the selfgravitating gas
for zero and nonzero cosmological constant $\Lambda$ 
while in section IV we give the physical 
picture of the gas and its particle density as a function of  $\Lambda$.
Section V contains the results of our Monte-Carlo simulations on the phase diagram, 
the collapse transition, the particle density and the properties of the collapsed
phase.

\section{The self-gravitating gas in the presence of the dark energy}

\subsection{Non-relativistic selfgravitating particles 
in the presence of the cosmological constant}

We recall some results about the self-gravitating gas in the presence of the 
cosmological constant \cite{gaslambda}. In the Newtonian limit the Einstein 
equations of general relativity become\cite{pee}
\begin{equation} \label{poiL}
\nabla^2 V = 4\pi \, G \, \rho -  8\pi \, G \, \Lambda \; .
\end {equation}
\noindent where $V$ stands for the gravitational potential, $\rho$ for the 
density of massive particles and $\Lambda$ for the cosmological constant.
Eq.(\ref{poiL}) determines the  weak and static gravitational field
produced by non relativistic matter in the presence of the cosmological 
constant. For zero cosmological constant we recover the usual Poisson equation,
as it must be. 

The Hamiltonian for such set of self-gravitating particles in the presence of 
the cosmological constant can be written in the center of mass frame 
as\cite{gaslambda}
\begin{equation} \label{hamil}
H = \sum_i \frac{{\vec p_i}^2}{2 \, m_i^2} -G \; \sum_{i<j} 
\frac{m_i \; m_j}{|{\vec q}_i-{\vec q}_j|}
- \frac{4\pi \, G \, \Lambda}{3} \; \sum_i m_i \; {\vec q}_i^{\,2}  \quad ,
\end {equation}
where $ m_i $ stands for the mass of the particle at the point $ {\vec
q}_i $ with momentum ${\vec p_i} $  and
$$
\sum_i  m_i \;  {\vec q}_i = 0 \; ,
$$
since we choose center of mass coordinates.

The cosmological constant contribution to the potential energy
grows negative for increasing values of the particle distances  ${\vec q_i}$
 to the center of mass.
Therefore, the gravitational effect of the  cosmological
constant on particles amount to push them outwards.
 It can be then said that the cosmological constant has an
{\bf anti-gravitational} effect. 

\subsection{Statistical mechanics of the self-gravitating gas with $ \Lambda
\neq 0$}

We present here the statistical mechanics 
of the self-gravitating gas in the presence of the cosmological
constant in the canonical ensemble.  For simplicity we shall consider  $N$ 
particles with identical mass $m$.

The Hamiltonian is given by eq.(\ref{hamil}) and therefore
the classical partition function of the gas is then,
$$
Z(T,N,V)=\frac{1}{N!} \int\ldots\int \prod_{l=1}^{N}
\frac{{\rm d}^3 {\vec  p_{l}} \; {\rm d}^3 {\vec q_{l}}}{(2 \pi)^3} \;
 \; e^{- \frac{H}{T} } \; \;.
$$
\noindent It is convenient to introduce the dimensionless coordinates
${\vec r_{l}}=\frac{{\vec q_{l}}}{L}$.
The momenta integrals are computed straightforwardly. Hence, the
partition function factorizes as  the partition function of 
a perfect gas times a coordinate integral.
 \begin{equation}\label{part}
Z=\frac{V^{N}}{N!} \left(\frac {m T}{2 \pi}\right)^{3 N/2}
 \; \int\ldots\int \prod_{l=1}^{N}{\rm d}^3{\vec r_{l}} \; \;
e^{\eta  \, u_{P} +\frac{2 \pi}{3} \, \xi \, u_{N}} \; ,
\end{equation}
where
\begin{equation} \label{upun}
u_{P} \equiv \frac {1}{N} \sum_{1 \leq i < j \leq N} 
\frac {1}{|{\vec r_{i}}-{\vec r_{j}}|} \quad , \quad
u_{N} \equiv \sum_{i=1}^{N}  r_{i}^2
 \; . 
\end {equation}
\noindent The dimensionless parameters which characterize the strength of 
self-gravity and dark energy are respectively,
\begin{equation} \label{etaxi}
\eta \equiv \frac{G\, m^2 \, N}{T \, L} \quad \mbox{and} \quad \xi   \equiv
2 \, \Lambda\, G\, m \, \frac{L^2}{T}  \; . 
\end {equation}
\noindent $ \eta $ can be interpreted as the ratio of the self-gravitating
energy  per particle to the kinetic energy per particle.

We define the ratio of dark energy and self-gravity 
\begin{equation} \label{x}
X \equiv \frac{\xi}{\eta}=\frac{2 \Lambda V}{m N} .
\end{equation}
It is the ratio between the dark energy contained inside the volume $V$
 and the mass of ordinary matter.

In the large $N$ limit the $3N$-uple integral in eq.(\ref{part}) can
be approximated by a functional integral over the density.
In the saddle point approximation we find that the density
\begin {equation} \label{phi} 
\rho({\bf x})=e^{\Phi({\bf x})} \; .
\end{equation}
obeys to the differential equation \cite{gaslambda}
\begin {equation} \label{Poisson}
\nabla^2 \Phi({\bf x})+4 \pi \; \left( \; \eta \; e^{\Phi({\bf x})}
-\xi \right)=0 \;.
\end{equation}
The density has to be normalized as
\begin {equation} \label{normalisation}
\int {\rm d}^3 {\bf x}\;\rho({\bf x})=1   \; \; .
\end{equation}
In hydrostatic equilibrium we obtain the same equation \cite{gaslambda}. 
Therefore, hydrostatics and mean field are equivalent in the $
N \to \infty $ limit.

\noindent  In the limiting case $X=0$ eq.(\ref{Poisson}) becomes the well known
 Lane-Emden equation in the absence of cosmological
constant (see refs. \cite{Emdsg,Chandrsg,Ebertsg,Bonnorsg,Antonovsg,Lyndsg,Katzsg,
Padmanabhansg,Saslawsg}). 
 
\subsection{The isothermal sphere with $ \Lambda
\neq 0$}

\begin{figure}[htbp]
\rotatebox{-90}{\epsfig{file=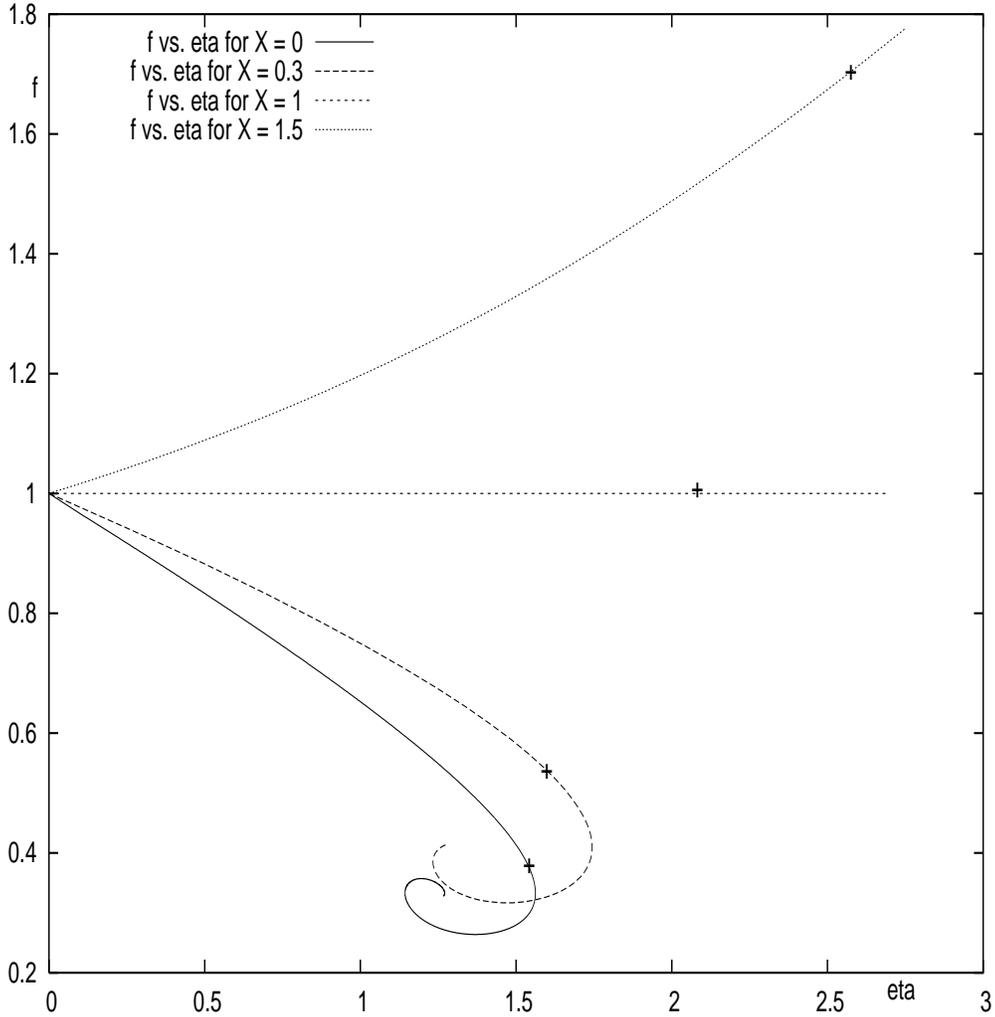,width=14cm,height=14cm}}
\caption{The density at the boundary  $f(X,\eta)$ versus $\eta$ for
$ X=0,0.3,1,1.5 $ by the mean field approach. The instability points
$ \eta_o(X) $ where the thermal compressibility diverges are pinpointed
 on the plot by a $+$.  
They are  $\eta_o(X=0)=1.510 \ldots$, $\eta_o(X=0.3)=1.63 \ldots$,
$ \eta_o(X=1)=2.04 \ldots$ and $\eta_o(X=1.5)=2.55\ldots$.
The gas is stable from $\eta=0$ till $\eta=\eta_o(X)$ in the 
canonical ensemble. } 
\label{ps}
\end{figure}

For spherically symmetric configurations the mean field equation 
(\ref{Poisson})
becomes an ordinary non-linear differential equation.
\begin {equation} \label{Poissonrad}
\frac{{\rm d}^2 \Phi}{{\rm d}R^2}+\frac{2}{R} \frac {{\rm d} \Phi}{{\rm
d}R} +4 \pi \; \left(\eta \; e^{\Phi(R)}-\xi \right) = 0 \; .
\end{equation}
\noindent 
 The various thermodynamic quantities are expressed in terms 
of their solutions. 
We work in a unit sphere, therefore the radial variable runs
in the interval $0 \leq R \leq R_{max} $,
$ R_{max} \equiv \left(\frac{3}{4 \pi}\right)^{\frac{1}{3}} $.
The density of particles  $\rho(R)$ has
to be normalized according to eq.(\ref {normalisation}).
Integrating eq.(\ref{Poissonrad}) from $R=0$ to $R=R_{max} $ yields,
\begin{equation} \label{emx}
\eta-\xi=-R_{max}^2 \; \Phi'(R_{max})
\end{equation}
\noindent Setting,
\begin{equation} \label{transf}
\Phi(R)=u(x)+\ln{\frac{\xi^{R}}{\eta^{R}}} \quad , \quad x=\sqrt
{3 \xi^{R}} \; \frac{R}{R_{max}} \; ,
\end{equation}
\noindent 
$\xi^{R}=\frac{\xi}{R_{max}}$ and $\eta^{R}=\frac{\eta}{R_{max}}$,
the saddle-point equation (\ref{Poissonrad}) simplifies as,
\begin{equation} \label{reducedeq}
\frac{{\rm d}^2 u}{{\rm d}x^2}+\frac{2}{x} \frac {{\rm d} u}{{\rm
d}x}+e^{u(x)}-1=0 \; .
\end{equation}
\noindent 
In order to have a regular solution at origin we have to impose
\begin{equation} \label{up0}
u'(0)=0 \; .
\end{equation}
\noindent
We find from eqs.(\ref{emx})-(\ref{transf}) for fixed values of $\eta$
and $\xi$,
\begin{equation} \label{emx2}
u'\left(\sqrt{3 \xi^{R}} \right)=-\frac{\eta^{R}-\xi^{R}}{\sqrt{3 \xi^{R}}}
 \; .
\end{equation}
Eqs.(\ref{up0}) and (\ref{emx2}) provide the boundary conditions for
the nonlinear ordinary differential equation (\ref{reducedeq}). In
particular, they impose the dependence of $u_0 \equiv u(0)$ on
$\eta^{R}$ and $\xi^{R}$. 

\noindent  We find from  eq.(\ref{reducedeq}) for small $x$ the 
expansion of  $u(x)$ in powers of $x$ with the result,
\begin{equation} \label{dl}
u(x)=u_0  + (1-e^{u_0}) \; \frac{x^2}{6}+e^{u_0} (e^{u_0}-1)   \;
\frac{x^4}{120} + {\cal O}(x^6)\; .
\end{equation}
\noindent where we imposed eq.(\ref{up0}). The value of $u_0$ follows
by imposing eq.(\ref{emx2}).

\noindent  Using eqs.(\ref{x}), (\ref{phi}) and (\ref {transf}) 
we can express the
density of particles in terms of the solution of eq.(\ref {reducedeq})
\begin{equation} \label{dens}
\rho(R)=X \; \exp\left[u\left(\sqrt
{3 X \eta^{R}} \; \frac{R}{R_{max}}\right)\right]\; .
\end{equation}
\noindent  We choose to express the physical quantities in terms of 
$\eta$ [eq.(\ref{etaxi})] and the ratio $X$ of dark energy to self-gravity 
[eq.(\ref{x})].
The local pressure obeys the ideal gas equation of state in local form 
$ P(R)=\frac{N \; T}{V} \; \rho(R) \; . $
Using eqs.(\ref{dens}) the pressure at the boundary is given by 
\begin{equation} \label{f}
P(R_{max})=\frac{N \; T}{V} \; X \; \exp\left[u\left(\sqrt {3
  X  \eta^{R} }\right)\right]\; \equiv \frac{N \; T}{V} \; f(X,\eta). 
\end{equation}
\noindent The contrast $ C(X,\eta) $
between the pressure at the center and at the boundary can be  written as, 
$$
C(X,\eta)\equiv\frac{P(0)}{P(R_{max})} \; =e^{u_0-u\left(\sqrt
{3 X \eta^{R}}\right)} \; .    
$$
\noindent We plot in fig. \ref{ps} the density at the boundary
versus $\eta$ for different values of $X$. 

\section{Stability of the self-gravitating gas}

\begin{figure}[htbp]
\rotatebox{-90}{\epsfig{file=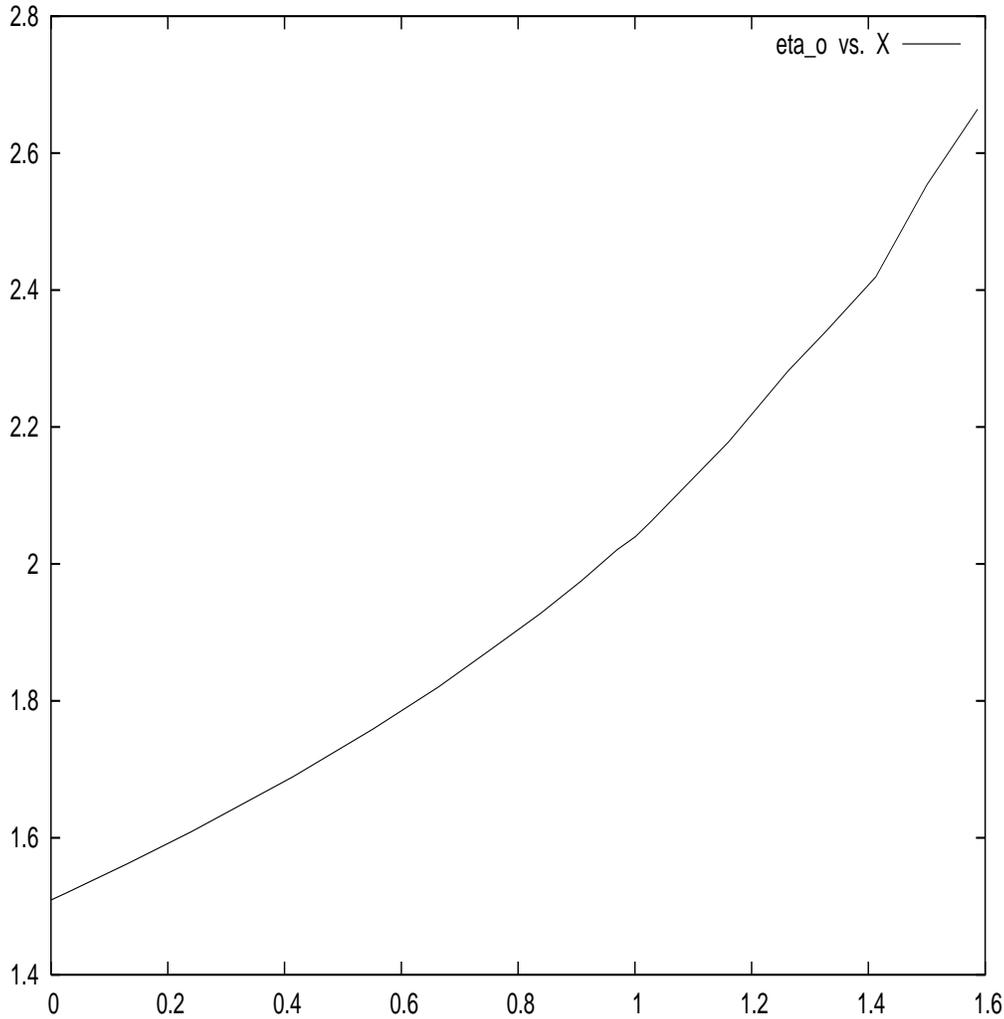,width=14cm,height=14cm}}
\caption{The value of the parameter $\eta$ at the Jeans instability: 
$\eta=\eta_o(X)$ versus $X$.}
\label{etao}
\end{figure}

\subsection{Stability for $\Lambda=0$.}

The self-gravitating
system  can be in one of two phases: gaseous or highly condensed. 
Mean field theory describes the gas phase and gives exactly the 
physical quantities in the thermodynamic limit.
The equilibrium density $\rho({\vec x})$ solution of the mean field
equation (\ref{Poisson}) 
minimizes the Helmholtz free energy in the {\bf canonical} ensemble.
This density is the most probable distribution which becomes absolutely 
certain in the infinite $N$ limit. All thermodynamics quantities follows 
from this density  $\rho({\vec x})$. This continuous  density $\rho({\vec x})$ 
solution of the mean field equation (\ref{Poisson}) fails to describe the 
condensed phase.

The condensed phase have been found by Monte-Carlo 
simulations for  $\Lambda=0$ \cite{gasn}. In the microcanonical ensemble the 
center collapses in a very dense core surrounded by a halo of particles. In 
the canonical ensemble all the particles collapse in a very dense body 
\cite{gasn}. 

The stability of the gaseous phase in the mean field
approach can be analyzed looking for the extrema of the dimensionless
parameter\cite{Lyndsg}
\begin{equation}
\label{zeta} 
\zeta=\frac{G^3 \; m^6 N^2 \; P}{T^4} \; .
\end{equation}
when the temperature $T$, the pressure $P$ and the number of particles $N$ are 
kept fixed in the canonical ensemble.  The gas phase becomes instable 
(for $\Lambda=0$) when $\zeta$ is maximal\cite{Lyndsg}. 

Using eqs.(\ref{etaxi}), (\ref{f}) and (\ref{zeta}) we find that
\begin{equation}\label{zetaet}
\zeta=\eta^3 \; f_0(\eta) \; .
\end{equation}
where $f_0(\eta) \equiv f(X=0,\eta)$ is the external density in absence of 
dark energy.

A better physical insight is obtained by looking to the the isothermal 
compressibility 
\begin{equation}
\label{defkt} 
K_T=-\frac{1}{V}  \; \left( \frac{\partial V}{\partial P}
\right)_{T} \; .
\end{equation}
\noindent Using eqs. (\ref{etaxi}) and (\ref{f}) we find that the 
dimensionless isothermal compressibility 
$\kappa_T \equiv \frac{NT}{V} \; K_T$ takes the form
\begin{equation} \label{kappat} 
\kappa_T = \frac{1}{f_0(\eta)+\frac{\eta  }{3} f_0^{'}(\eta)} \; .
\end{equation}
We see from eqs.(\ref{zetaet}) and (\ref{kappat}) that
$$
\kappa_T = \frac{3 \; \eta^2}{\frac{d\zeta}{d\eta}} \; .
$$
Hence, $\kappa_T$ {\bf diverges} at the extrema of $ \zeta $.

The isothermal compressibility $\kappa_T$ is positive from $\eta=0$ 
till $\eta=\eta_o=1.510 \ldots$. At this point $\kappa_T$ as well
as the specific heat at constant pressure diverge and change
their signs\cite{Ebertsg,gasn}. Moreover, at this point 
the speed of sound at the center of the sphere becomes imaginary\cite{gas2}. 
Therefore, small density 
fluctuations will grow exponentially in time instead of exhibiting oscillatory 
propagation. Such a behaviour leads to the Jeans instability and collapse\cite{Jeans}.
Monte-Carlo simulations confirmed the presence of this instability at 
 $\eta=\eta_o=1.510 \ldots$ in the canonical ensemble for $\Lambda=0$ 
\cite{gasn}.

\subsection{Jeans instability for $\Lambda \neq0$}

We now compute the dimensionless isothermal compressibility for the 
self-gravitating gas with $\Lambda \neq 0$.
Introducing the dimensionless parameter
\begin{equation} \label{alpha}
\alpha=\frac{2 \; G^3 \; m^5 \; N^2 \Lambda}{T^3} \; = \eta^3 \; X , 
\end{equation}
the dimensionless isothermal compressibility $ \kappa_T $ takes the form:
\begin{equation} \label{kappatlamb} 
\kappa_T = \frac{1}{f(\eta,\alpha)+\frac{\eta  }{3} \; 
\left(\frac{\partial f}{\partial \eta} \right)  (\eta,\alpha)}   \; .
\end{equation}
\noindent The Jeans instability happens when the isothermal compressibility 
diverges. This happens for $\eta=\eta_o(X)$ verifying:
$$
f(\eta_o,\alpha)+\frac{\eta_o  }{3} \; 
\left(\frac{\partial f}{\partial \eta} \right)  (\eta_o,\alpha) = 0 \; .
$$
Thus, for a fixed $\alpha$, $\eta_o$ is the value of $\eta$ which
maximizes the quantity 
\begin{equation} \label{zetax} 
\zeta \equiv \eta^3 \; f(\eta,\alpha) =\frac{G^3 \; m^6 \; N^2 \; P}{T^4} \; .
\end{equation}
\noindent As for the self-gravitating gas with
$\Lambda=0, \;  \zeta$ is maximal when the specific heat at constant
pressure diverges and changes its sign.
We plot $\eta_o$ versus $X$ in fig. \ref{etao} using eq.(\ref{alpha}).
Monte Carlo simulations confirm that the onset of the Jeans collapse happens 
at  $ \eta = \eta_o $. 

We can write $ \eta $ [eq.(\ref{etaxi})] as
$$
\eta = \frac{G\, m \, M}{T \, L}
$$
where $ M $ is the total mass of the gas. Therefore, the collapse
condition $ \eta \geq  \eta_o(X) $ can be written as
$$
M \geq M_0 \; \eta_o(X) \quad {\mbox where} \quad M_0 \equiv
 \frac{L \, T}{G \, m} \; .
$$
The gas collapses when its mass takes a larger value than the critical one $ M_J(X) 
\equiv M_0 \; \eta_o(X) $. We can consider $ M_J(X) $ as the generalization
of the Jeans mass in the presence of the cosmological constant. Notice that
the presence of $ \Lambda $ {\bf increases } the Jeans mass with respect the
$ \Lambda = 0 $ case [see fig. \ref{etao}].

\section{Physical Picture}

\subsection{Behaviour of the gas phase with X}

The effects of self-gravitation and dark energy go in opposite directions. 
Self-gravitating 
forces are attractive, while dark energy produces repulsion. If the ratio 
of dark energy and self-gravity is 
$X<1$ [defined by eq.(\ref{x})], self-gravitation dominates over dark energy. 
If $X=1$ the effect of self-gravitation is exactly compensated by the
 dark energy. If $X>1$ dark energy dominates over self-gravitation. 
We illustrate these three cases plotting $f(X,\eta)$ versus $\eta$ for 
$X=0.3,1,1.5$ in fig. \ref{ps}

\begin{itemize}

\item {\bf first case: $X<1$: self-gravity dominates}

We illustrate this case choosing $X=0.3$. We see in fig. \ref{ps} that 
$f(X,\eta)$  versus $\eta$ exhibits a form  analogous to 
$f(X=0,\eta)$ versus $\eta$ in absence of dark energy. $f(X,\eta)$ has two 
Riemann sheets as a function of $\eta$.  
In the first sheet $f(X,\eta)$ monotonically decreases with
$\eta$ for fixed $ X < 1 $.

\item {\bf second case: $X=1$: exact compensation}

The effect of self-gravitation is compensated by the effect of dark energy.
 We see in fig. \ref{ps} that $f(X=1,\eta)=1$.
An exactly homogeneous sphere $\rho(R)=1$ is a solution (see below).
In such special case the self-gravitating gas behaves as a perfect gas
(with $PV=NT$ everywhere). 

\item {\bf third case: $X>1$: dark energy dominates}

We illustrate this case choosing $X=1.5$. We see in fig. \ref{ps} that 
$f(X=1.5,\eta)$ is an increasing function of $\eta$ for fixed $ X > 1 $.

\end{itemize}

\noindent The Jeans instability happens in the three cases. The value of
$ \eta = \eta_o(X)$ where the gas collapses increases with $X$ 
(fig. \ref{etao}). The gas collapses for $ X>0$ at a higher density
$\frac{N}{V^{\frac{1}{3}}}$ for a given temperature or for a lower temperature
for a given  density $\frac{N}{V^{\frac{1}{3}}}$ than for $X=0$. That is, the 
domain of stability of the gas increases for increasing $X$. The dark energy
has an anti-gravity effect that disfavours the collapse pushing the particles
toward the boundary of the sphere. Notice that the point of Jeans instability 
is in the first Riemann sheet of the $\eta$-plane for $0 \leq X<1$.
  
\subsection{The particle  density $\rho(R)$}

\begin{figure}[htbp]
\rotatebox{-90}{\epsfig{file=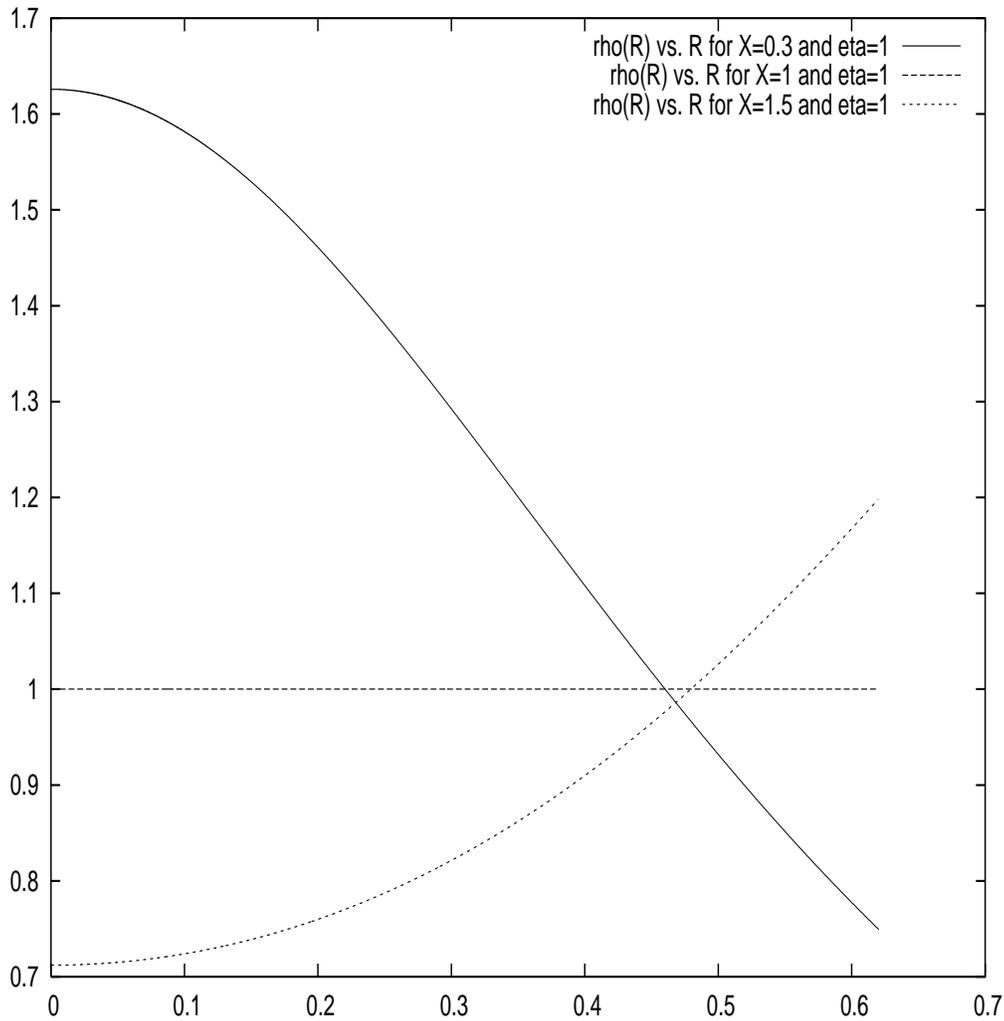,width=14cm,height=14cm}}
\caption{The particle densities $\rho(R)$ vs. the radial coordinate $R$ 
for $X=0.3$ and $\eta = 1$ (decreasing), $X=1$ and $\eta = 1$ (homogeneous) 
and finally $X=1.5$ and $\eta =1$ (increasing) computed by the mean field 
approach}
\label{rho}
\end{figure}

We study the density $\rho(R)$  as a function of the radial coordinate $R$. The
density $\rho(R)$ of the self-gravitating gas in absence of dark
energy ($X=0$) is always a {\bf decreasing} function of  $R$. This follows 
from the attractive character of gravitation (see for example ref. 
\cite{gasn,gas2}).

The $R$-dependence of the density
$\rho(R)$ of a self-gravitating gas in the presence of dark energy is more
involved because the dark energy opposes to the attraction by gravity. 
The functions $\rho(R)$ and $u(x)$ have the same qualitative dependence in
$R$ since they are related by exponentiation [see eq. (\ref{dens})].

The behaviour  at the center of the sphere ($R=0$) is
governed by the sign of $u_0$ since $u'(0)=0$ [eq. (\ref{up0})] and we find
from eq.(\ref{dl}) that $ u''(0)= \frac{1}{3} \! 
\left(1-e^{u_0} \right)$. Hence,
$$
\mbox{sign}[ u''(0) ] = -\mbox{sign}[ u_0 ] \; .
$$
Therefore, $\rho(R)$ decreases (increases) at $R=0$ if $u_0>0$ ($u_0<0$).

The behaviour at the boundary of the sphere
($R=R_{max}$) is governed by the sign of $\eta-\xi$ since according to
eq.(\ref{emx2})
$$
\mbox{sign}[ u'(R_{max}) ] = -\mbox{sign}[ \eta^{R}-\xi^{R} ] =  
\mbox{sign}(X-1) \; .
$$
Therefore, $\rho(R)$ decreases (increases) at $R_{max}$ if $X<1$ ($X>1$).

\medskip

The particle density exhibits for the stable solutions (namely, for
$\kappa_T > 0$) one of the following behaviours illustrated in fig. \ref{rho}

\begin{itemize}

\item{\bf decreasing}: $X<1$ and $u_0>0$. The density $\rho(R)$ decreases
 from the center of the sphere till the boundary. We plot in fig. \ref{rho}
the density $\rho(R)$ versus $R$ for $X=0.3$ and $\eta=1$. The self-gravitation
dominates over dark energy.

\item{\bf increasing}: $X>1$ and $u_0<0$. The density $\rho(R)$
  increases from the center of the sphere till the boundary. 
We plot in fig. \ref{rho}
the density $\rho(R)$ versus $R$ for $X=1.5$ and $\eta=1$. The dark energy
dominates over self-gravitation.

\item{\bf homogeneous}: $X=1$ and $u_0=0$. The density is homogeneous:
$\rho(R)=1$. The effect of self-gravity of particles is exactly compensated by
the effect of dark energy. We plot in fig. \ref{rho}
the density $\rho(R)$ versus $R$ for $X=1$ and $\eta=1$.
 \end{itemize}

Eq.(\ref{Poisson}) for $ X=1 $ is  similar to the equation that describes
in two space dimensions (multi)-vortices in the
Ginsburg-Landau or Higgs model in the limit between superconductivity
of type I and II \cite{vortex}. However, for the vortex case one has $
 \xi = \eta < 0 $ since like charges repel each other in electrodynamics 
while masses attract each other in gravity. This opposite nature of the
forces is responsible of the fact that eq.(\ref{Poisson}) for $ X=1 $ only
has trivial stable solutions while a host of non-trivial solutions
appear in the vortex case\cite{vortex}.

\section{Monte Carlo calculations}

\subsection{The Metropolis algorithm}

The standard Metropolis algorithm \cite{MC} was first applied to 
self-gravitating gas in absence of dark energy in ref.\cite{gasn}.
We apply this standard Metropolis  to the self-gravitating gas
in presence of dark energy. We perform it in a volume $V$ in the 
canonical ensemble at temperature $T$. We compute in this way the external 
pressure, the energy, the average density, the average particle distance and 
the average squared particle distance as function
of $\eta$ for a given value of  the ratio of dark energy 
and self-gravity $X \equiv \frac{\xi}{\eta}=\frac{2 \Lambda V}{m N}$.

We implement the Metropolis algorithm in the following way. We start from a 
random distribution of $N$ particles in the chosen volume. We update such
configuration choosing a particle at random and 
changing at random its position. We then compare the energies of the former and 
the new configurations.
We use the standard Metropolis test to choose between the new and the former
configurations. The energy of the configurations
are calculated performing the exact sums as in eq.(\ref{upun}). We use as 
statistical weight for the Metropolis algorithm in the canonical ensemble,
$$
e^{\eta  \, u_{P} +\frac{2 \pi}{3} \, \xi \, u_{N}} \; ,
$$
[see eqs.(\ref{part}) and (\ref{upun})]. 
The number of particles go up to $2000$. 

We introduce a  short distance cutoff in the Newtons' potential 
[see eq.(\ref{upun})] 
$$
u_{P} \equiv \frac {1}{N} \sum_{1 \leq i < j \leq N} 
\frac {1}{|{\vec r_{i}}-{\vec r_{j}}|_A}  \; ,
$$
with
\begin{eqnarray}
|{\vec r_{i}}-{\vec r_{j}}|_A 
&=&|{\vec r_{i}}-{\vec r_{j}}| \quad , \quad
  \mbox{for} \quad  |{\vec r_{i}}-{\vec r_{j}}| > A \nonumber\\
&=&A  \quad , \quad \mbox{for} \quad  |{\vec r_{i}}-{\vec r_{j}}| 
< A  \; , \nonumber
\end{eqnarray}
\noindent where $A \ll  V^{\frac{1}{3}}$ is the short distance cut-off.
The presence of the  short distance cut-off  prevents the collapse
(here unphysical) of the self-gravitating gas (for more details see 
\cite{gasn}).

\subsection{Influence of the geometry}

We apply the Metropolis algorithm to the self-gravitating gas in presence 
of dark energy in a sphere and in a cube.

\noindent For $X=0$ (absence of dark energy) the Monte Carlo results
 in a sphere and in a cube with identical volume 
give practically the same results \cite{gasn}.
The geometry does not influence the physics of the system.

\noindent However, for $X>0$ (presence of dark energy) the Monte Carlo results
in a  sphere and in a cube turn to be different. This effect can be traced
to the repulsive character of the dark energy that pushes the particles towards
the boundary. As a result, the physical quantities becomes sensible to the 
geometry. We thus compared the  mean field results with spherical symmetry 
from the previous sections and ref.\cite{gaslambda}  with Monte Carlo
calculations performed on a sphere.

\subsection{Phase diagrams}

\begin{figure}[htbp]
\rotatebox{-90}{\epsfig{file=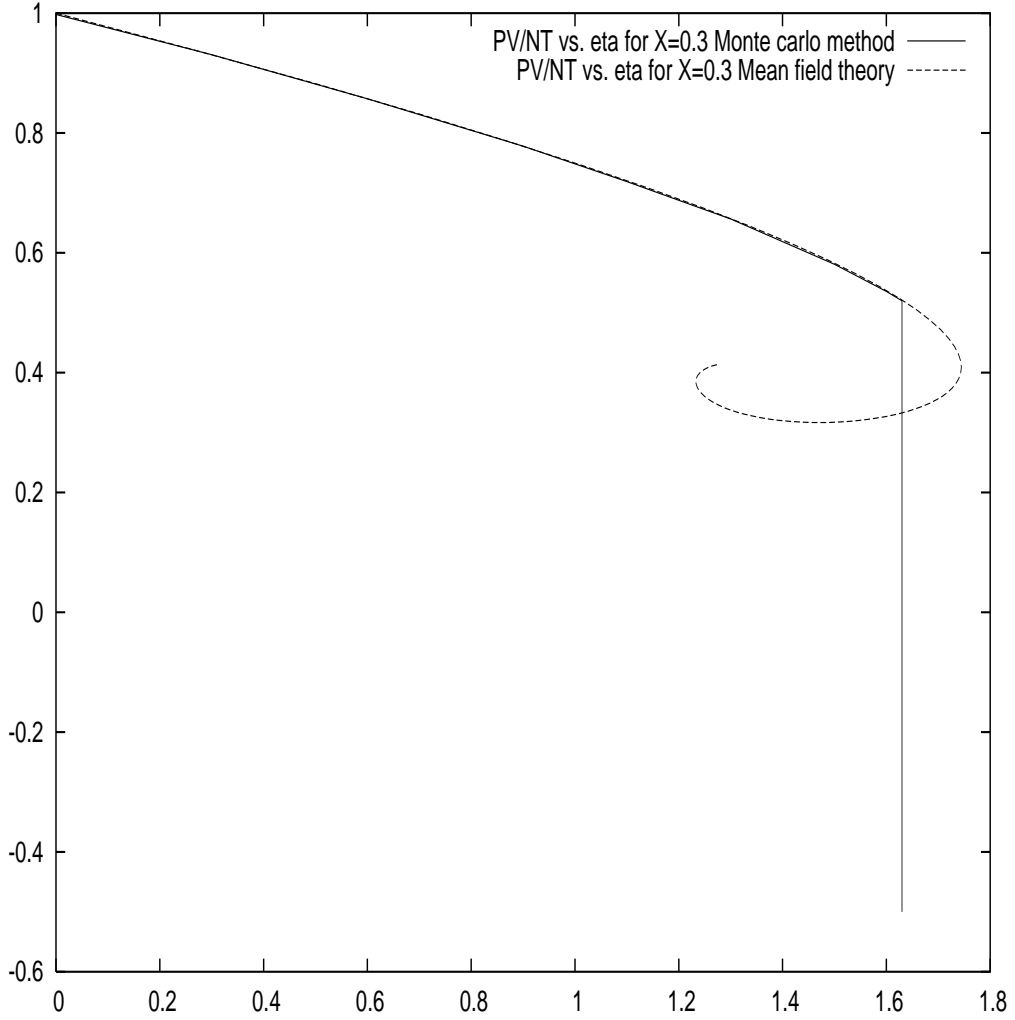,width=14cm,height=14cm}}
\caption{The density at the boundary  $f_X(\eta)=\frac{PV}{NT}$ versus $\eta$ 
for $X=0.3$ by Monte Carlo methods and mean field approach. The collapse value
of the gas phase is $\eta_T(X=0.3)=1.63\ldots$}
\label{fmc0p3}
\end{figure}

\begin{figure}[htbp]
\rotatebox{-90}{\epsfig{file=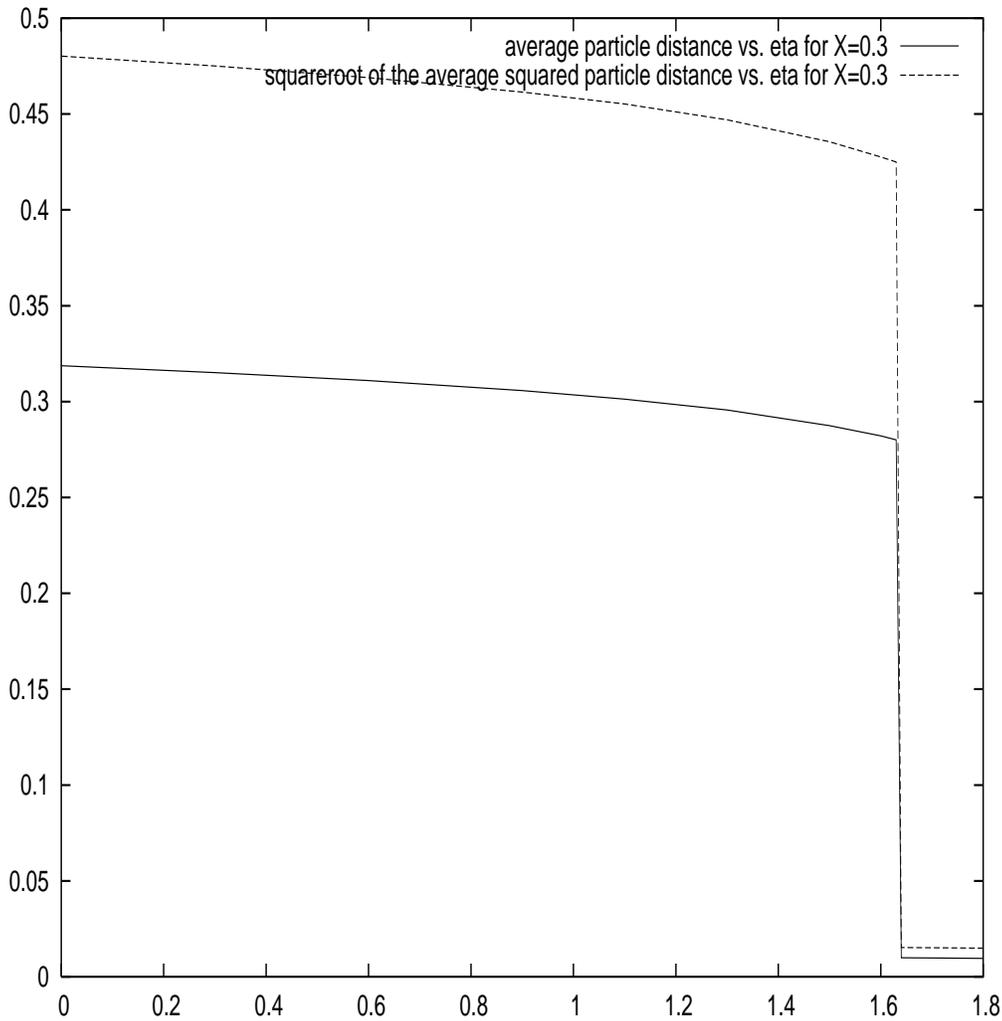,width=14cm,height=14cm}}
\caption{The average particle distance and the average squared particle 
distance  versus $\eta$ for
$X=0.3$ by Monte Carlo methods. They exhibit a
sharp decrease at the collapse value $\eta=\eta_T(X=0.3)=1.63\ldots$.}
\label{rmc}
\end{figure}

\begin{figure}[htbp]
\rotatebox{-90}{\epsfig{file=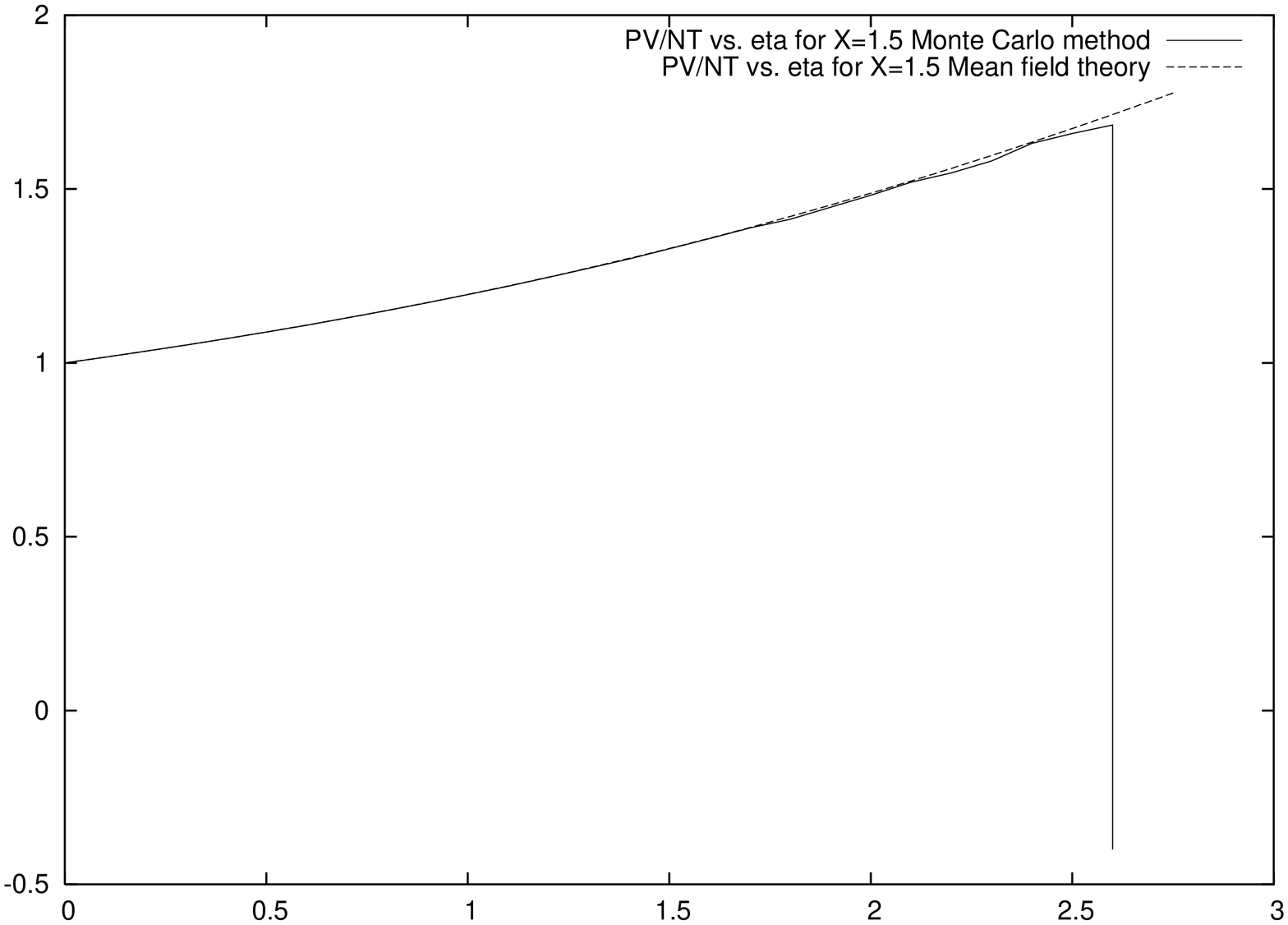,width=14cm,height=14cm}}
\caption{The density at the boundary  $f_X(\eta)=\frac{PV}{NT}$ versus $\eta$ 
for $X=1.5$ by Monte Carlo methods and mean field approach. The collapse value
of the gas phase is $\eta_T(X=1.5)=2.6\ldots$}
\label{fmc1p5}
\end{figure}

We recall that our self-gravitating  system exhibits three different 
behaviours according the value of $X$, 
the relative rate of dark energy and self-gravity.
In all  three cases two different phases show up:
for $\eta < \eta_T(X)$ we have a non perfect gas and for $\eta > \eta_T(X)$
it is a very condensed system with negative pressure. The transition between the
two phases is very sharp.

\begin{itemize}

\item For $X<1$ the self-gravity dominates over dark energy.
We plot the external density $f(X,\eta)= \frac{PV}{NT}$ versus $\eta$
for $X=0.3$ in fig. \ref{fmc0p3}. In  the gas phase
$\frac{PV}{NT}$ monotonically decreases with $\eta$ from $\eta=0$ 
till the collapse point $\eta=\eta_T(X)$.

\noindent The average distance between particles $<|{\vec r}_i - {\vec r}_j|>$
and the average squared distance between particles
$\sqrt{<|{\vec r}_i - {\vec r}_j|^2>}$ monotonically decrease with
$\eta$ (fig. \ref{rmc}). When the gas collapses at $\eta_T$, 
$<|{\vec r}_i - {\vec r}_j|>$ and $\sqrt{<|{\vec r}_i - {\vec r}_j|^2>}$ 
exhibit a sharp drop.

\item For $X=1$ the effect of dark energy exactly compensates the effect of
self-gravity. We have $\frac{PV}{NT}=1$ in the gas phase. 
The system behaves like a perfect gas  from $\eta=0$ 
till the collapse point $\eta=\eta_T(X)$.

\item For $X>1$ the dark energy dominates over self-gravity.
We plot in fig. \ref{fmc1p5} the external density 
$f(X,\eta)=\frac{PV}{NT}$ versus $\eta$ for $X=1.5$.
In the gas phase $\frac{PV}{NT}$ monotonically increases with $\eta$ 
from $\eta=0$ till the collapse point $\eta=\eta_T(X)$.

\end{itemize}

We find that the Monte Carlo calculations in the sphere 
accurately reproduce the mean field results for $\frac{PV}{NT}$ in the
gas phase with spherical symmetry. 
We then consider a cube and a sphere of identical volume. The
Monte Carlo calculations turn to give different results in these two cases:
a) The external pressure is larger in the cube than in the  sphere. 
b) The gas phase is more stable in the cube. c) The collapse value $\eta_T(X)$ 
in the cube is larger than its value for the same $X$ in the sphere.  

\subsection{The collapse point}

The phase transition happens in the Monte Carlo simulations
at $\eta=\eta_T(X)$. It must be compared with 
the point of Jeans instability $\eta=\eta_o(X)$ according to mean field
[see III.C] where the isothermal compressibility diverges and changes its sign.
We see in  that $\eta_T(X)$ is {\bf very close} of $\eta_o(X)$. 
They are probably the same point. For $X=0.3$ we have $\eta_o(X)=1.63 \ldots$
and  $\eta_T(X)=1.63 \ldots$.
 For $X=1.5$ we have $\eta_o(X)=2.55 \ldots$
and  $\eta_T(X)= 2.6\ldots$.
We conclude that the collapse of the system observed in the Monte Carlo 
simulations is due to the Jeans instability when the isothermal compressibility
diverges and is well described by the mean field approach.

For $\eta \gtrsim \eta_T(X)$ we see that the gaseous phase is metastable.
Before collapsing the system stay for a long Monte Carlo time in the 
metastable gaseous  phase in the Monte Carlo computations. 
The  Monte Carlo time for collapse increases with
increasing value of $X$. This is due to the repulsive effect of dark energy.
The gas phase is less  unstable in the presence of dark energy. 

\subsection{Average distribution of particles}

We illustrate the two behaviours of the self-gravitating gas in the presence
of the cosmological constant  plotting
the average density as a function of two Cartesian coordinates, the third
coordinate being fixed. For simplicity we depict the densities in the cube. 

\begin{figure}[htbp]
\rotatebox{-90}{\epsfig{file=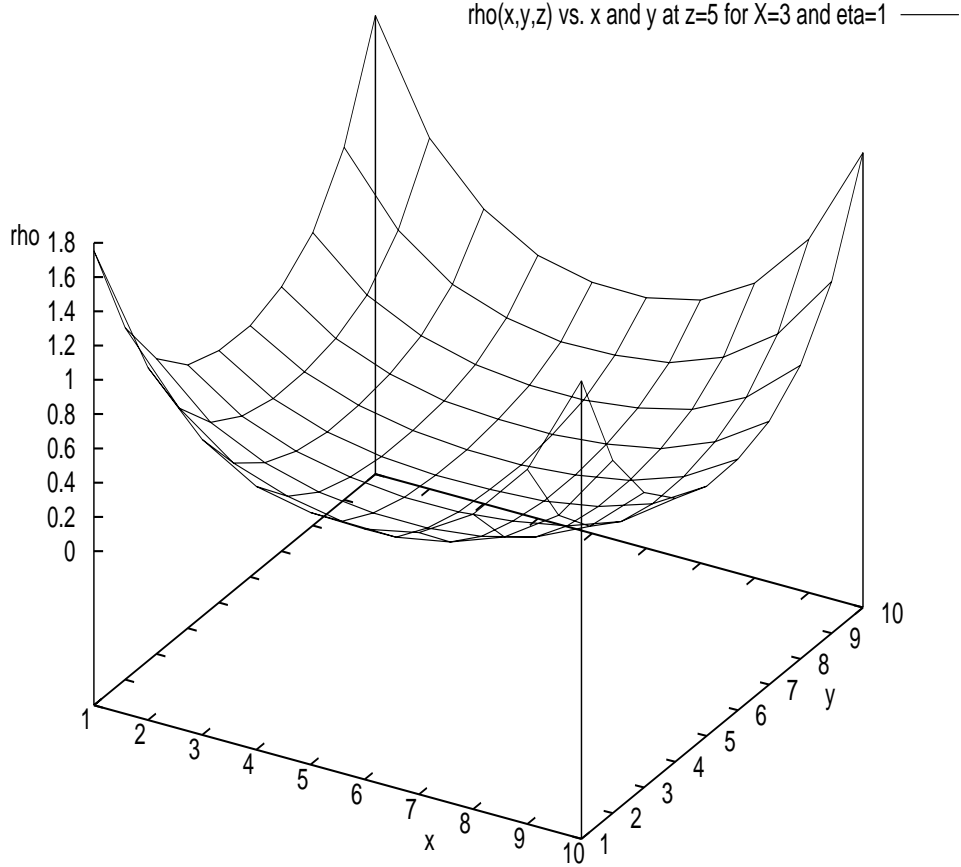,width=14cm,height=14cm}}
\caption{The local density of the gas $\rho(x,y,z)$ versus
the Cartesian coordinates $x$ and $y$ at $z=5$ in a cube of size $9$
for $X=3$ and $\eta=1$ from Monte Carlo simulations. 
The density is larger on the boundary of the cube than at the center. }
\label{densinc}
\end{figure}

\noindent For $X>1$ we see that the density is larger on the boundary that
at the center of the cube (fig. \ref{densinc}). The dark energy dominates over 
the self-gravity and pushes the particles toward the boundary of the cube.

\begin{figure}[htbp]
\rotatebox{-90}{\epsfig{file=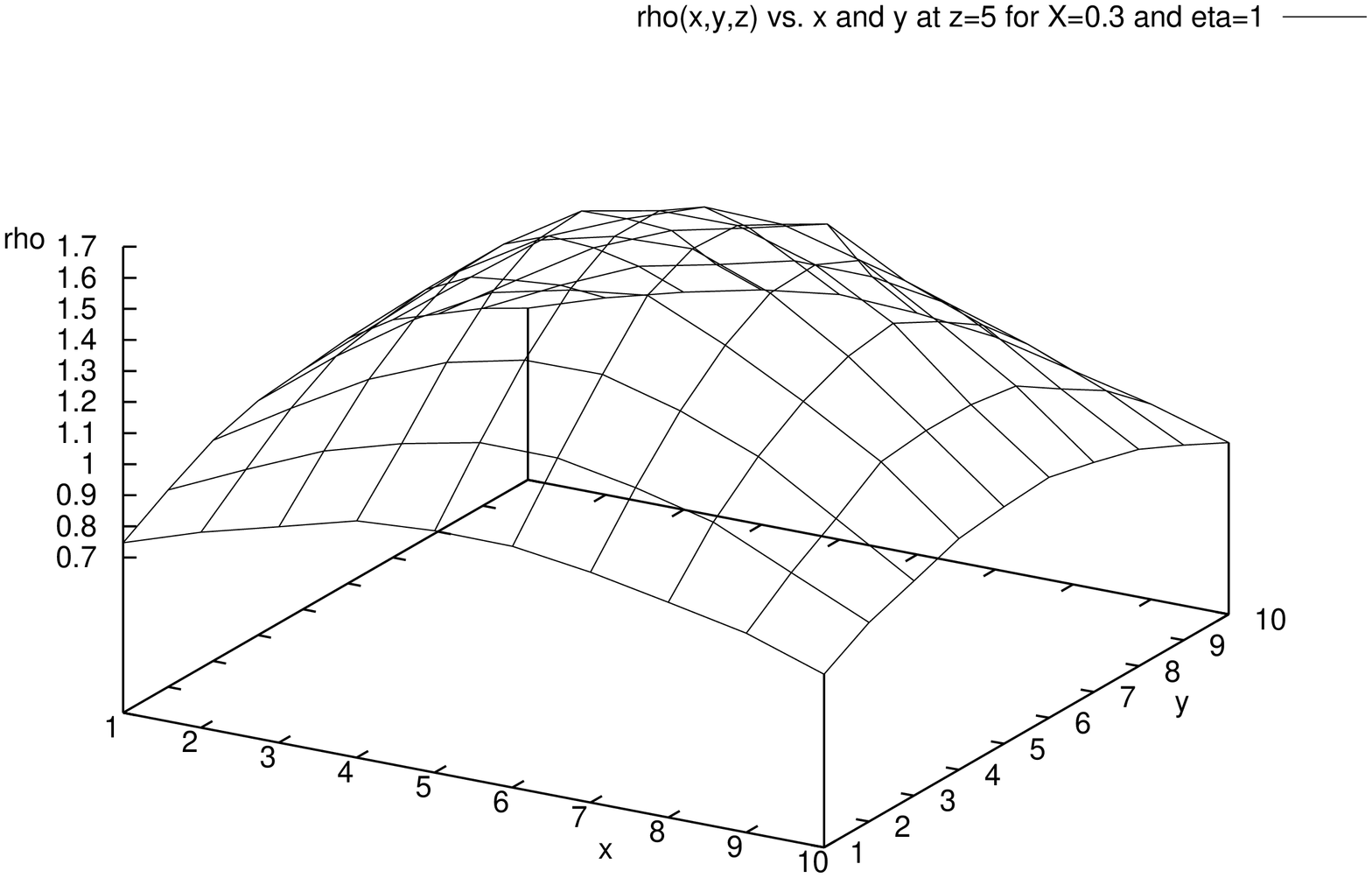,width=14cm,height=14cm}}
\caption{The local density of the gas $\rho(x,y,z)$ versus
the Cartesian coordinates $x$ and $y$ at $z=5$ in a cube of size $9$
for $X=0.3$ and $\eta=1$ from Monte Carlo simulations. 
The density is larger at the center of the cube than on the boundary.}
\label{densdec}
\end{figure}

\noindent For $X<1$ we see that the density is larger at the center than on
the boundary of the cube (fig. \ref{densdec}).
The self-gravity dominates over dark energy and attracts the particles to the 
center of the cube.

\begin{figure}[htbp]
\rotatebox{-90}{\epsfig{file=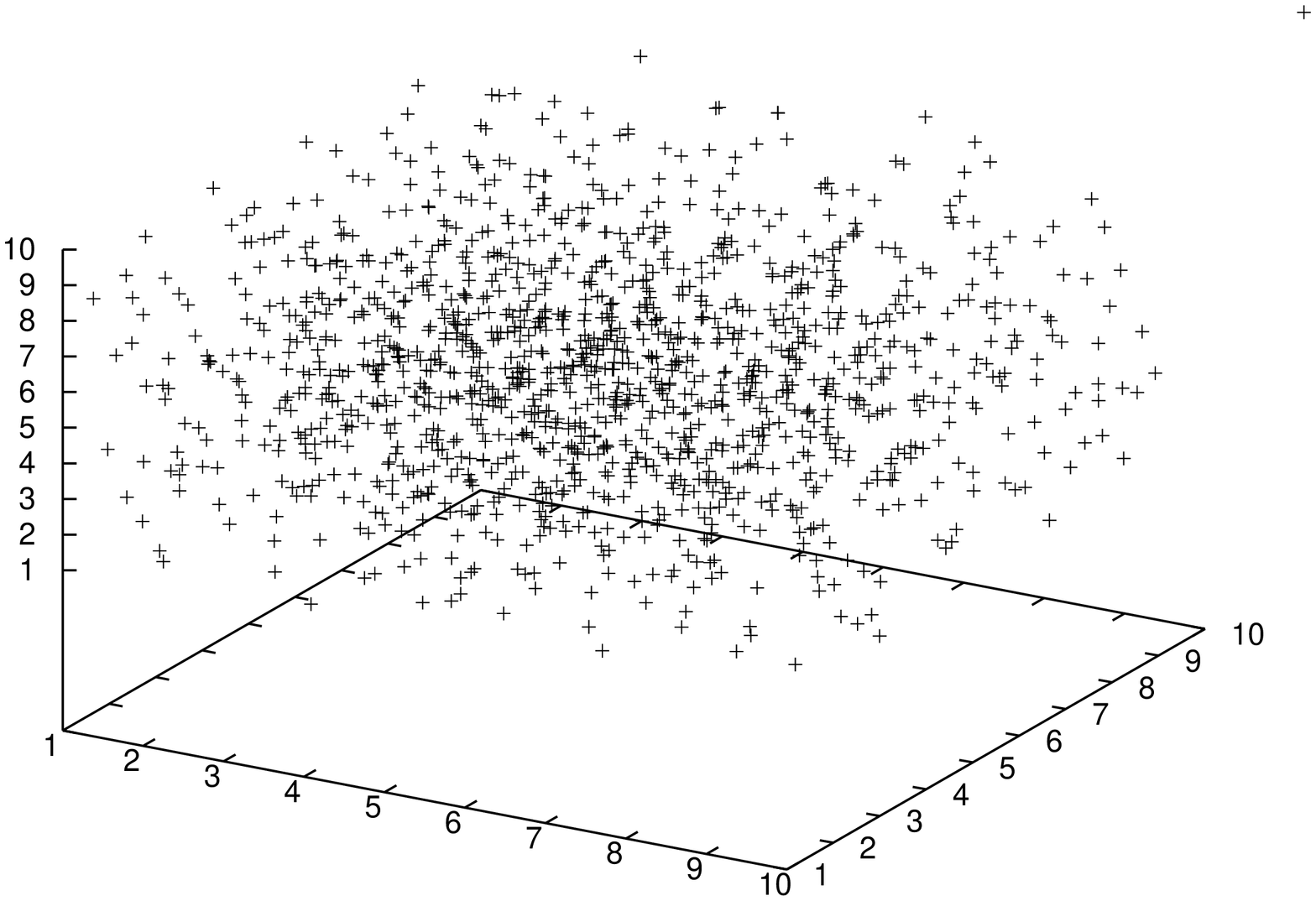,width=14cm,height=14cm}}
\caption{Average particle distribution in the gaseous phase from Monte 
Carlo simulations in a cubic volume for $X=0.3$, $\eta=1.6$ and $N=2000$. 
A $+$  denotes one particle.}
\label{profgas}
\end{figure}

\begin{figure}[htbp]
\rotatebox{-90}{\epsfig{file=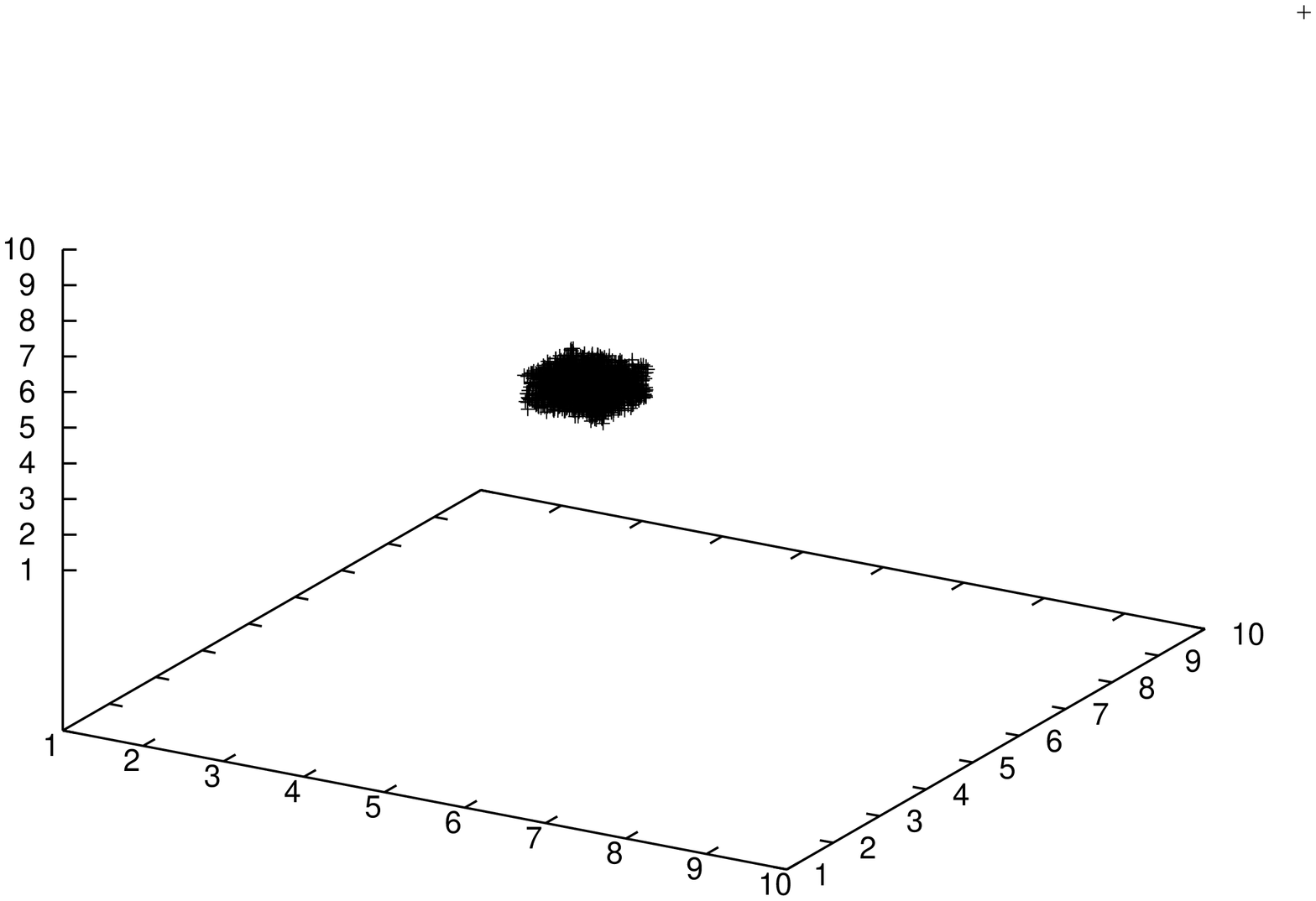,width=14cm,height=14cm}}
\caption{Average particle distribution in the collapsed phase from Monte 
Carlo simulations in a cubic volume for $X=0.3$, $\eta=1.8$ and $N=2000$. 
A $+$  denotes one particle.}
\label{profcoll}
\end{figure}

We illustrate now the two phases (gaseous and condensed) 
plotting the average particle distribution in a cube.
Fig. \ref{profgas} and fig. \ref{profcoll}  depict the average particle 
distribution
from Monte Carlo calculations with $1000$ particles for $X=0.3$ at both sides
of the collapse point. Fig. \ref{profgas} corresponds
to the gaseous phase and fig. \ref{profcoll} to the collapsed phase. 

We find that the Monte Carlo simulations (describing thermal equilibrium)
are much more efficient than the $N$-body simulations integrating Newton's 
equations of motion. [Indeed, the integration of Newton's equations
provides much more information than thermal equilibrium investigations].
Actually, a few hundreds of particles are enough to get
quite accurate results in the Monte Carlo simulations (except near the collapse
points). Moreover, the Monte Carlo results turns to be in excellent
agreement with the mean field calculations up to corrections of the 
order $ \frac1{N} $. 

\subsection{The condensed phase}

In the condensed phase all the particles collapse in a very dense body
(see fig. \ref{profcoll}). The self-gravity contribution of the potential 
energy dominates overwhelmingly  the dark energy contribution in the condensed 
phase.  Using eqs. (\ref{hamil}), (\ref{etaxi}) and (\ref{x}) 
and the virial theorem \cite{gaslambda}
we obtain that the external pressure is expressed as
$$
f_X(\eta)=\frac{PV}{NT}=1-\frac{1}{3} 
\left< \frac{1}{|{\vec r}_i-{\vec r}_j|} \right> \; \eta \; .
$$
The Monte Carlo results indicate that 
$\left< \frac{1}{|{\vec r}_i-{\vec r}_j|} \right> \simeq 50$.
Thus in this condensed phase the external pressure can be
approximated by
$$
f_X(\eta)=\frac{PV}{NT}=1-K \eta \; .
$$
where $K \simeq 16$. 

Since $ f(\eta) $ has a jump at the transition, the Gibbs free energy
is discontinuous and we have a phase transition of the {\bf zeroth} order
as in the absence of cosmological constant \cite{gasn}.

\section{Discussion and Conclusions}

The behaviour of the self-gravitating gas is significantly influenced 
by the cosmological constant (or not) depending on the value of the ratio
$X$ defined by eq.(\ref{x}),
$$
X = \frac{2 \, \Lambda \; V}{m \; N} = 2 \;
\left(\frac{\mbox{cosmological \; constant}}{\mbox{mass}}\right)_V = 2 \;
\frac{\Lambda}{\rho_V}   
$$
where $\rho_V \equiv N \; m /V $ is the average mass density in the
volume $V$ and where the subscript $\left( \right)_V$ indicates mass and 
cosmological constant inside the volume $V$.

$X$ takes small values for astrophysical objects except for clusters
of galaxies where $ X{galaxy clusters} \sim 0.01 $ and especially for the universe 
as a whole where  $ X{universe} \sim 4 $. 
The smallness of  $X$ stems from the tiny value of the present cosmological constant 
$\Lambda = 0.663 \; 10^{-29}$g/cm$^3$\cite{lam}. The particle mass density is
much larger than $\Lambda$ in all situations except for the universe as a
whole. 

However, the non-relativistic and equilibrium treatment does not
apply for the universe as a whole.
Certainly, the galaxy distribution can be considered as a non-relativistic 
self-gravitating gas in the presence of the cosmological constant. 
But the gas of galaxies is certainly not yet at thermal equilibrium today. 
Perhaps, thermal equilibrium may be reached in some billions of years.
Incidentally, at thermal
equilibrium the particle density is regular at the center of the distribution
in the whole range of $ X $. Namely, there is no cusp  at the center
in a situation of thermal equilibrium.

\end{document}